Submitted to the *Journal of Double Star Observations* March23,2023# Automated Speckle Interferometry of Known Binaries

(With WDS 12274-2843 B 228 as an example)

Nick Hardy[1], Leon Bewersdorff[2], David Rowe[3], Russell Genet[4], Rick Wasson[5], James Armstrong[6], Scott Dixon[7], Mark Harris[8], Tom Smith[9], Rachel Freed[4], Paul McCudden[10], S. Stephen Rajkumar Inbanathan[11], Marie Davis[4], Christopher Giavarini[4], Ronald Snyder[4], Roger Wholly[4], Maaike Calvin[11], Sumner Cotton[11], Julia Carter[11], Mario Terrazas[11], Shane Christopher R.[12], Arun Kumar A.[12], Sithara Naskath H.[12],
 and Mariam Ronald Rabin A.[12]

[1] OurSky, Enschede, Netherlands
[2] Fairborn Institute, Aachen, Germany
[3] PlaneWave Instruments, Torrance, California, USA
[4] Eastern Arizona College, Payson, Arizona, USA
[5] Orange County Astronomers, Temecula, Calif
[6] University of Hawaii, Institute for Astronomy, Maui, USA
[7] San Diego Astronomy Association, San Diego, California, USA
[8] Oorsprong Science, Roswell, City, Georgia, USA
[9] Dark Ridge Observatory, Weed, New Mexico, USA
[11] Colorado Mountain College, Steamboat Springs, Colorado, USA
[10] American College, Madurai, India## Abstract

Astronomers have been measuring the separations and position angles between the two components of binary stars since William Herschel began his observations in 1781. In 1970, Anton Labeyrie pioneered a method, speckle interferometry, that overcomes the usual resolution limits induced by atmospheric turbulence by taking hundreds or thousands of short exposures and reducing them in Fourier space. Our 2022 automation of speckle interferometry allowed us to use a fully robotic 1.0-meter PlaneWave Instruments telescope, located at the El Sauce Observatory in the Atacama Desert of Chile, to obtain observations of many known binaries with established orbits. The long-term objective of these observations is to establish the precision, accuracy, and limitations of this telescope's automated speckle interferometry measurements. This paper provides an early overview of the Known Binaries Project and provide example results on a small-separation (~0.27″) binary, WDS 12274-2843 B 228.## 1. Introduction

*Binary Stars*

Starting in 1781, Herschel (1782) observed pairs of stars (double stars) that appeared close together in the sky. He assumed, logically, that the brighter star of the pair was nearer to the Earth and the fainter one further away. He expected to see the bright star shift back and forth with respect to the faint one as the Earth orbited the Sun (parallax motion). Simple trigonometry could then determine the distance to the bright star. While Herschel didn't detect any parallactic motion, he did detect a curved relative motion between the two stars and correctly concluded that they were actually very close to each other—not just coincidentally aligned along our line-of-site—and were rotating synchronously around the pair's common center of gravity (Herschel 1803). He named these gravitationally bound stellar pairs *binaries*.

By repeatedly measuring the separation between a binary's two stars and their position angle with respect to celestial north, the binary's elliptical orbit can eventually be determined. Knowing the binary's period and the distance to the binary, Kepler's Third Law provides the combined mass of the two stars which can be divided between the two components using several methods (Argyle 2012, Heintz 1971). Establishing the orbits of binaries through measuring their changing separations and position angles is the most accurate way to determine stellar masses. Knowing stellar mass is the key to understanding stellar evolution.

Astronomers have been measuring and publishing the position angles and separations of double stars since the early 1780s (Herschel 1782). Since 1964, the *Washington Double Star Catalog* (WDS), maintained by the U.S. Naval Observatory, has maintained the consolidated record of all published observations (http://www.astro.gsu.edu/wds/). While most of the over 100,000 double stars in the WDS are mere optical doubles (chance alignments), some are binary pairs. Orbits have been published for ~ 3000 of these known binaries in the *Sixth Catalog of Orbits of Visual Binary Stars* (Matson et al. 2023).

*Speckle Interferometry*

It is difficult to accurately establish orbits (and hence stellar masses) if the periods are centuries or millennia; it is much easier if they are just a few years or decades. However, to have short periods, the two binary components must be physically close to each other. Furthermore, they must be relatively near to Earth, or they cannot be resolved as two separate stars (Genet 2015). Although larger aperture telescopes in theory have greater spatial resolution than smaller telescopes, in practice the resolution of Earth-bound telescopes was limited to about an arcsecond—even at the best sites—due to atmospheric turbulence until a workaround was discovered. Anton Labeyrie's (1970) paper, 'Attainment of diffraction limited resolution in large telescopes by Fourier analyzing speckle patterns in star images' opened the door to high resolution observations. Speckle interferometry began in earnest with Harold McAlister's observations of many sub-one-arcsecond binaries on the 2.1- and 4.0-meter telescopes at Kitt Peak National Observatory in Arizona in the early 1980s (McAlister 1985), as well as on the 3.9-meter Anglo-Australian telescope at Siding Spring Observatory in Australia.

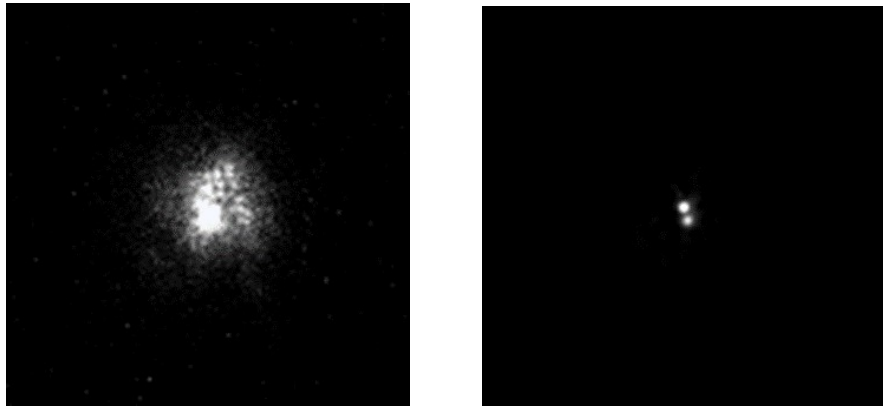

*Figure 1: Left: A highly magnified speckle interferometry image, one of 300 short-exposures (10 msec) of our example binary, WDS 12274-2843 B228, taken automatically with the PW1000 speckle system in 2.4" seeing. The short exposures freeze out atmospheric turbulence smearing, but there are multiple light paths to the camera that create the scrambled speckle image. Mathematically complex and computer intensive bispectrum analysis is used to recreate the image, largely removing the atmospheric effects. Right: The reconstructed bispectrum image of the binary star approaches the diffraction limited resolution of the telescope. The two very close stars are clearly seen, although their separation is only 0.26". Both images have the same scale,*

*8.3″ (128 pixels) square.*

Although fully automated photometric measurements were routinely made by the mid-1980s (Genet et al. 1987), speckle interferometry is one of the last astronomical observational techniques to be fully automated. An early attempt at automation (Teiche 2015) could not be sustained. Recently, a software plugin, developed by two of us (Nick Hardy and Leon Bewersdorff) for NINA (Nighttime Imaging 'N' Astronomy), was successfully used on 1.0- and 0.6-meter PlaneWave Instruments robotic telescopes located at the automated El Sauce Observatory in the Atacama Desert of Chile. Fully automated observations were obtained in the late 2022 and early 2023 for several research efforts including this Known Binaries Project. Hardy and Bewersdorff, loaded target lists, monitored the operation of the telescopes, and placed the observational results in a Google drive for access by the researchers. Several the coauthors of this paper helped solve the many problems that arose during these initial months of automated observations.

*Known Binaries Project Overview*

Observing known binaries with established orbits adds additional points to these orbits leading, eventually, to more accurate orbits and hence more accurate stellar mass estimates. Also, since the projected future separations and position angles are known with some certainty (ephemerides in the *Sixth Orbit Catalog,* Matson et al. 2023), these 'known' future positions can be compared with what is observed to help characterize the performance and accuracy of an observational system. While there are errors in the projected orbital positions of known binaries (and these errors cannot be separated from observational errors), they can be minimized by observing systems with higher grade (more accurate) orbits and observing a sizeable number of known binaries to search for observational biases (inaccuracies).

Binaries with known (published) orbits are reported in the *Sixth Catalog of Orbits of Visual Binary Stars* (Mateson et al. 2023). A spreadsheet extract from this catalog (McCudden et al. 2022) was used to make up lists of known binaries potentially observable by the 1.0-meter telescope in Chile starting in the Fall of 2022. The fully automated speckle interferometry observations covered a wide range of binary parameters. Most of the observed binaries had component magnitudes between 6 and 13, separations 0.2 and 1.00″, periods between 4 and 500 years, and grades of 1 to 3 (with 1 being highest quality orbits and 5 the lowest). Some of the binaries were just observed on a single night, while others were observed on multiple nights. A few were observed twice in the same night.

*Published Student Research*

When it comes to sports, everyone gets it; you must play the game to understand what it is all about. It would be ludicrous to teach basketball by just practicing layups, free throws, and jump shots without playing any games. While science lectures and labs are important, there is simply no substitute for students participating early on in published research to supercharge their STEM careers.

Astronomy is considered by many to be a gateway to science and technology, thanks to its ability to inspire curiosity in everyone irrespective of age or culture. The Astronomy Research Seminars have been producing published student-team papers since 2007. Student teams manage their own research, obtain original data from a robotic telescope, write a team paper, obtain an external review, and submit their paper for publication. Students are supported by experienced researchers and the Institute for Student Astronomical Research. A National Science Foundation grant evaluated and expanded these workshops to many other schools. Over the years, some 200 published papers have been coauthored by some 700 high school and community college students. Thanks to their research experience, many workshop graduates have obtained admission to a college of their choice, often with a scholarship.

The Seminars originally used astrometric eyepieces on small telescopes supplied by local amateurs to observe rather bright, widely separated, long-period binaries. Starting in 2014, the seminars utilized CCD cameras on remote robotic telescopes to observe fainter, more closely separated, somewhat shorter period binaries. These CCD observations were limited, however, by atmospheric fluctuations (seeing) to separations of several arcseconds. In this Known Binaries Project, seminar students have been able to use automated speckle interferometry on a remote 1-meter robotic telescope to obtain observations of truly short-period binaries with separations approaching just 0.2″.

*Goals of the Known Binaries Project*

The overall goals of the Known Binaries Project are to:

(1) Explore the performance limits of the automated 1.0-meter telescope.

(2) Establish the precision and accuracy of the automated speckle interferometry observations.

(3) Add our observations to each binary's growing number of observations, eventually resulting in
improved orbital solutions.

Given the small separation of some of the most interesting targets, we have been:

- Exploring the human-induced variance in the final reduction process (which involves human judgment in the centering of diaphragms on the binary's primary and secondary stars following the final bispectrum image reconstruction process.

- Trying, algorithmically, to fit two Gaussian curves that approximated the point spread function (*i.e.,* the seeing) to the somewhat merged image of some of the close stars.

- Considering how reducing the observed bispectrum image sizes, from 512x512 to 128x128 or, alternatively, applying spatial filtering in Fourier space might improve the results (an effect first noted by one of us—Mark Harris).

Finally, because some of the initial full-frame long exposures used to determine the plate scale and camera angle were missing, we determined if variance of the plate scale and camera angle on many targets observed on the same and multiple nights were sufficiently steady so mean or bracketed values could be used in place of missing or failed plate solutions.

## 2. Instrumentation and Automation

Observations for the Known Binaries Project were obtained with a PlaneWave 1-meter PW1000 Corrected Dall-Kirkham optical telescope. The telescope is located at the El Sauce Observatory in Chile at an altitude of 1600 meters. One of the telescope's two Nasmyth instrument ports is dedicated to speckle interferometry. This port contains a TeleVue 2x Powermate Barlow lens, a QHY filter wheel, and a QHY600MM-Pro CMOS camera. The camera is ideally suited for speckle interferometry, thanks to its fast image readout, low readout noise of 1.68e-, excellent quantum efficiency in the R-band, and negligible dark current when cooled to -10°C. The 9-slot filter wheel includes Sloan g', r', z' and i' photometric band filters.

With the Barlow lens installed, the telescope has an effective focal length of 12038 mm and subsequent instrumental image scale of 0.065″/pixel. The AltAz-mounted telescope uses a field de-rotator to keep the image at a constant orientation. The mount's tracking accuracy error is less than 1" over 10 minutes. The telescope is automatically refocusing every 30 minutes. Seeing at El Sauce is typically under 2".

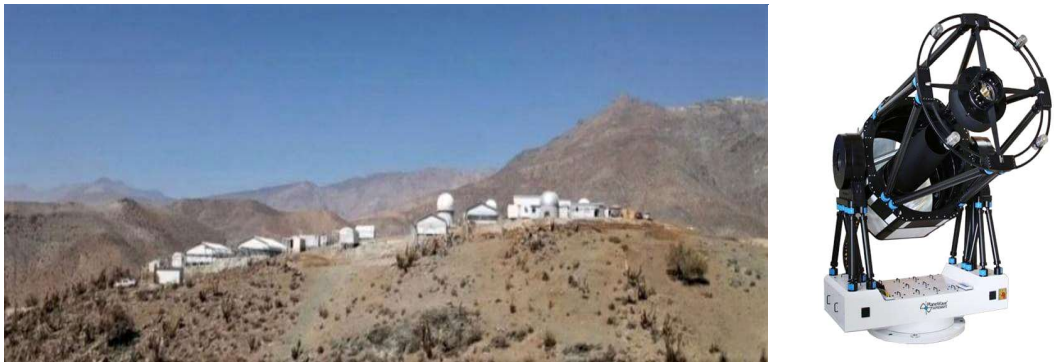

*Figure 2: Left: the remote, fully automatic El Sauce Observatory. Solar panels provide the power. Right: the 1.0-meter Planewave Instruments robotic telescope. It is housed in one of the roll-off roof enclosures.*

The operation of the telescope is fully automated through use of NINA (Nighttime Imaging 'N' Astronomy), the PlaneWave PWI4 telescope control software, and a new NINA Speckle Interferometry plugin developed by two of the authors (Nick Hardy and Leon Bewersdorff). Initially, a target list is loaded into a list within the NINA sequence. Each target is automatically assigned a best imaging time within the night by the plugin based on the time at which the targets are at their highest altitude. Targets also need to be above 50° degrees altitude and at least 20° away from the moon. Unobservable targets within the target list are automatically rejected for that night.

At any given time in the night, the target with the nearest imaging time is chosen and loaded into the sequencer with its predefined sequence template. The telescope waits for the target's best imaging time to arrive if it has not passed yet. The telescope then slews to the target's coordinates and takes a 60 second full-frame exposure which, after obtaining a plate solution, allows a region of interest (ROI) to be precisely centered on the target. This full-frame image is saved along with the target data to determine the precise plate rotation and image scale in later reduction. A 512 by 512 pixel (33.3" x 33.3") ROI is then centered on the target's x,y coordinates in the full-frame image. If plate-solving fails for any reason (sometimes the target is in a sparse starfield), then a bright star that is closest to the center of the full-frame image will be chosen on which to center the ROI. The appropriate filter is selected and the requested number of exposures of the ROI are taken by the camera at a high frame rate using the camera's high-gain video readout mode. Exposure times are calculated automatically by the plugin if not otherwise specified by the user.

The plugin's automatic calculation involves using simulations of the signal-to-noise ratio (SNR) of the target in the camera's images required to detect the secondary star at a certain SNR value in the final reduced image. This estimate is relatively consistent across different telescopes, with a difference in exposure time that is required to reach the necessary SNR. The lower the exposure time, the closer a theoretical ideal exposure time is approached in which precisely one Fried diameter is frozen. Thus, the exposure time from the automatic calculation is the minimum required to still see the secondary star at a certain SNR, while approaching the ideal exposure time as closely as possible. Directly calculating and using the theoretical ideal exposure time is a more complex process that requires an empirical model of the smearing of the images caused by the atmosphere. However, such models are not readily available, vary from night to night, and typically lead to the underexposure of most targets.

The exposure calculation approach, which aims to get as close to the ideal exposure time as possible, is similar to what an operator would manually do, decreasing the exposure time while ensuring the star still is reasonably exposed. Specifics on the calculation process will follow in a future paper. The resulting exposure times are typically 10- to 100-ms, although longer exposures are sometimes manually requested or calculated for faint systems even if some seeing induced smearing of the results may occur.

The SIMBAD (2023) astronomical database is then queried for a Smithsonian Astrophysical Observatory (SAO) star brighter than magnitude 12 (but no brighter than magnitude 8) that is within a distance 12° of the target to use as a reference star. The telescope slews to the nearest chosen reference star and repeats the entire process, capturing 300 images of an equally sized ROI before moving onto the next target.

At the conclusion of the night, the observations are automatically arranged into .fits cubes and Fast Fourier Transformed. The images and partially reduced data are immediately uploaded to the cloud via a local installation of the Google Drive software after they are saved for online access and further processing.

Various issues were overcome during the development of this plugin for fully automated speckle interferometry. The camera's video mode was no longer supported in NINA, as NINA had previously shifted focus to Deep Sky imaging. The mode had to be re-added by Hardy. He also added support for a plate-solving software option for NINA after Rowe had created a new version. The automatic target selection in the image was also incrementally refined. Initially it required a successful plate-solve before the ROI could be placed. Later, automatic selection of the brightest star was added, since the target star is often, but not always, the brightest one in the field of view.

Autofocuses in sparse star fields with the small FOV oftentimes resulted in a poor focus. They were made more reliable by querying SIMBAD for a star cluster within 25° of zenith, then slewing to it before running the autofocus routine, after which it would slew back to the target. Having numerous brighter stars in the image greatly improved focusing.

## 3. WDS 12274-2843 B 228

Many dozens of close known binaries have now been observed automatically in this program. They are being analyzed and papers are being written. WDS 12274-2843 B 228 was chosen for this early paper as an example of what we can obtain on a small separation (approaching 0.2″) system.

*Previous Observations*

The previous observations of WDS 12274-2843 B228, spanning nearly 100 years, are shown graphically in the WDS orbit plot in Figure 3 and quantitatively in Table 1.

The first observation of WDS 12274-2843 B228 in 1926—and discovery credit—go to the famous Dutch double star astronomer W. H. van den Bos (WDS symbol B) of Leiden Observatory, using the 26-inch Grubb refractor at the Union Observatory in Johannesburg, South Africa (van den Bos 1928). Van den Bos discovered more than 2895 double stars, made more than 71,000 measurements, and became director of Union Observatory in 1941.

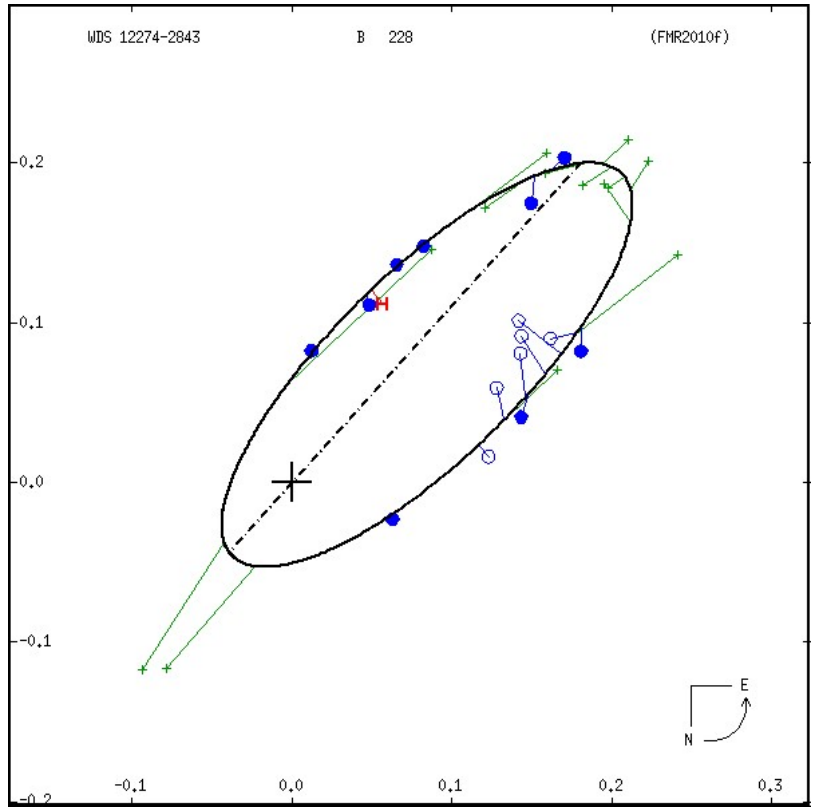

*Figure 3: Orbital plot of WDS 12274-2843 B228 taken from the Sixth Catalog of Orbits of Visual Binary Stars (Matson et al. 2023).*

Table 1 offers a good example of the evolution of observing technology. Early techniques were eyepiece micrometer (Ma and Mb, "+" symbols in Figure 3) and eyepiece interferometer (J, open blue circles). More than two orbits of the 44.5-year period have been completed since discovery, but the more accurate speckle interferometry technique (solid blue circles) was not applied to this binary until 1983 by the CHARA team (Hartkopf, et al. 2000), continuing with professional collaborators. Our observations appear to be the first speckle observations made with a telescope smaller than 4 meters.

*Table 1: Past observations of WDS 12274-2843 B228 from the WDS data base (supplied by the U.S. Naval Observatory). Column headings are date of observation, position angle (θ) in degrees, separation (ρ) in arcseconds, aperture of the telescope in meters, number (#) of nights averaged for the observation, the reference to the published observation, and the observing technique.*

| Date | θ | ρ | Aperture | #Nights | Reference | Technique |
|---|---|---|---|---|---|---|
| 1926.77 | 133.7 | 0.27 | 0.7 | 3 | B_1928b | Ma |
| 1929.99 | 132.0 | 0.30 | 0.7 | 4 | B__1930a | Ma |
| 1933.68 | 135.7 | 0.26 | 0.7 | 4 | B__1937b | Ma |
| 1936.24 | 140.8 | 0.25 | 0.7 | 4 | B__1937b | Ma |
| 1940.44 | 145.0 | 0.21 | 0.6 | 2 | Smw1951 | Ma |
| 1949.38 | 149.0 | 0.17 | 0.7 | 1 | B__1956a | Ma |
| 1952.32 | 141.4 | 0.15 | 0.7 | 1 | B__1956a | Ma |
| 1953.37 | :146. | :0.14 | 0.7 | 1 | B__1956a | Ma |
| 1959.33 | q113.1 | 0.18 | 0.7 | 3 | B__1960a | Ma |
| 1959.50 | 97.4 | 0.124 | 0.7 | 2 | Fin1960b | J |
| 1960.51 | 114.8 | 0.141 | 0.7 | 3 | Fin1961 | J |

| | | | | | | | |
|---|---|---|---|---|---|---|---|
| 1961.51 | 119.4 | . | 0.164 | 0.7 | 4 | Fin1962a | J |
| 1962.50 | 122.5 | . | 0.170 | 0.7 | 3 | Fin1963a | J |
| 1963.485 | 125.5 | . | 0.174 | 0.7 | 1 | Fin1964a | J |
| 1964.35 | 120.5 | . | 0.28 | 0.7 | 4 | B__1965a | Ma |
| 1964.505 | 119.1 | . | 0.185 | 0.7 | 3 | Fin1965a | J |
| 1976.131 | 132.9 | . | 0.27 | 1.5 | 3 | Wor1978 | Mb |
| 1979.20 | 135.6 | . | 0.30 | 0.9 | 3 | Hei1980a | Mb |
| 1983.4295 | 140.0 | . | 0.265 | 3.8 | 1 | Hrt2000a | Sc |
| 1984.3770 | 139.4 | . | 0.230 | 3.8 | 1 | McA1987b | Sc |
| 1987.306 | 142.3 | . | 0.26 | 0.4 | 5 | Sca1991a | Ma |
| 1989.3052 | 150.9 | . | 0.169 | 4.0 | 1 | McA1990 | Sc |
| 1990.3397 | 154.3 | . | 0.151 | 4.0 | 1 | Hrt1993 | Sc |
| 1991.25 | 153. | | 0.125 | 0.3 | 1 | HIP1997a | Hh |
| 1991.3910 | :156.5 | | :0.121 | 4.0 | 1 | Hrt1993 | Sc |
| 1993.0920 | 171.6 | | 0.083 | 4.0 | 1 | Hrt2000a | Sc |
| 2001.0801 | 69.8 | | 0.067 | 4.0 | 1 | Msn2009 | Su |
| 2006.1941 | 105.9 | | 0.149 | 4.0 | 1 | Msn2009 | Su |
| 2009.2585 | 114.4 | | 0.1986 | 4.1 | 1 | Tok2010 | S |
| 2017.4291 | 126.2 | | 0.2740 | 4.1 | 2 | Tok2018c | St |
| 2021.2451 | 130.2 | | 0.2873 | 4.1 | 2 | Tok2022f | St |

*Our Observations*

Speckle observations of WDS 12274-2843 B228 were obtained on the nights of 5, 8 and 9 March 2023, and are ongoing. Shown below in Figure 4 is the Plate Solution (Rowe & Genet 2015), for the full-frame area around WDS 12274-2843 B228. PlateSolve is part of the software suite supporting PlaneWave telescopes, but the PS3.79 version has been specially developed to deal with the small fields-of-view inherent in speckle interferometry, and PS3.80 has been integrated for autonomous observations into the NINA plug-in. These versions use both the UCAC4 and Gaia DR2 databases, with stars as faint as G ~18, so there are usually adequate stars to achieve a highly accurate plate solution.

The image in Figure 4 is an unguided full frame, 60-second exposure with the Sloan r′ filter. Because of the high magnification (focal length 12 meters) the field is small, just 10.3 x 6.9 arc-min.

Even if the initial telescope slew is not perfectly centered on the target, NINA uses the plate solution to place the ROI on the target coordinates. The speckle ROI (512x512 pixels = 33″) is shown in Figure 4 as the small yellow box centered on the overexposed target star B228. It is almost never necessary to waste time moving the telescope again or taking a second plate solve image.

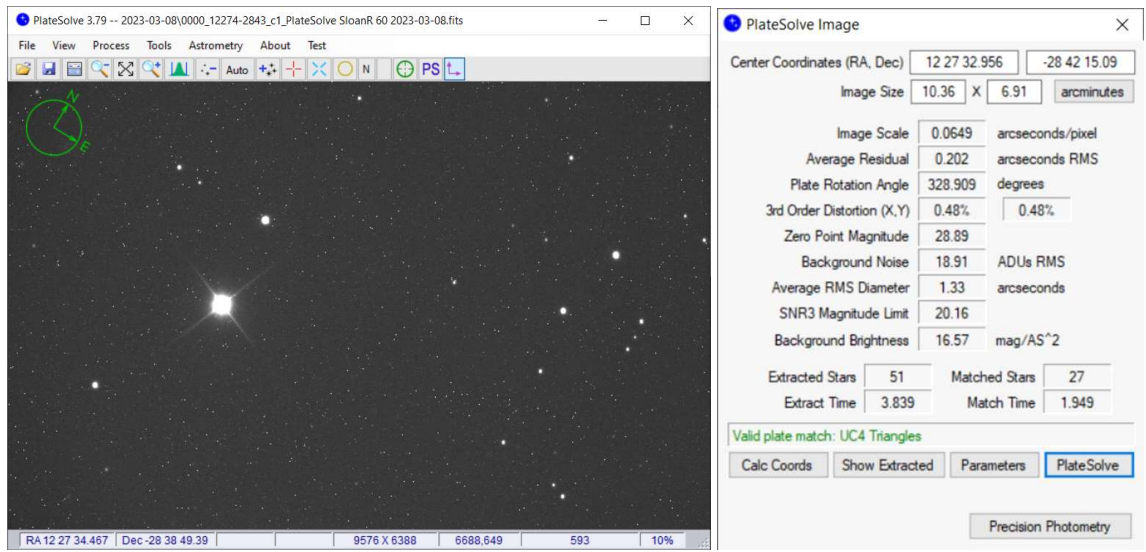

*Figure 4: Plate solution for WDS 12274-2843 B228 obtained 8 December 2023. The image scale was 0.0649"/pixel, while the camera angle was 328.91°. The solution matched 27 stars from the UCAC4 Catalog with an RMS residual of only 0.2". The seeing (RMS FWHM Diameter) at that time was 1.33", about average for the excellent Atacama Desert site. The bright sky background, 16.57 mag/arc-sec$^2$ (only one night from full moon), has minimal impact on speckle interferometry because of high magnification (f/12), allowing this telescope to be fully utilized on all clear nights.*

The Speckle Tool Box (STB) (Rowe and Genet 2015) is used for Bispectrum analysis of speckle images. STB is the product of years of development and is a unique software package for Windows that makes the powerful bispectrum method of image analysis freely available.

Several calibration tools are available in the current version, STB1.16, but minimal preparation is required for speckle processing (although added processing may be helpful to suppress noise for very faint binary stars). Bispectrum processing consists of re-formatting the ROI raw images as FITS cubes, running a triple correlation analysis of each image in the Fourier domain, and writing a .BSP1 file containing all those cross-correlations. Although this is a very computationally intense process, the user simply activates it by choosing tools from a drop-down menu.

The final steps of bispectrum analysis in STB are still done manually—they require judgment for several reasons. To scale the Fourier domain properly, inputs are required for telescope aperture, filter center wavelength, image scale, and desired enhancement (power) of reference star deconvolution. Phase recovery is an iterative process that begins with an autocorrelation—not a real image, but a Fourier transform of the power spectrum. The autocorrelation image has two equal lobes 180 degrees apart representing the secondary star; it can be measured for astrometry if the observer knows which lobe represents the real secondary star. Beginning with the autocorrelation, transition to the bispectrum image of the true stars is an iterative process, requiring judgment on when adequate convergence is achieved. Brightness adjustments and choice of Fourier filtering to suppress background noise are also used to obtain a pleasing final image.

Figure 5 below illustrates the procedures and results of recovering phase information in STB bispectrum analysis (i.e., reconstructing an image as though it had been taken with a plane wavefront). This image and measurement of B228 is one of many bispectrum analyses currently in work by students on the Close Known Binaries project, using data from the 1-meter telescope.

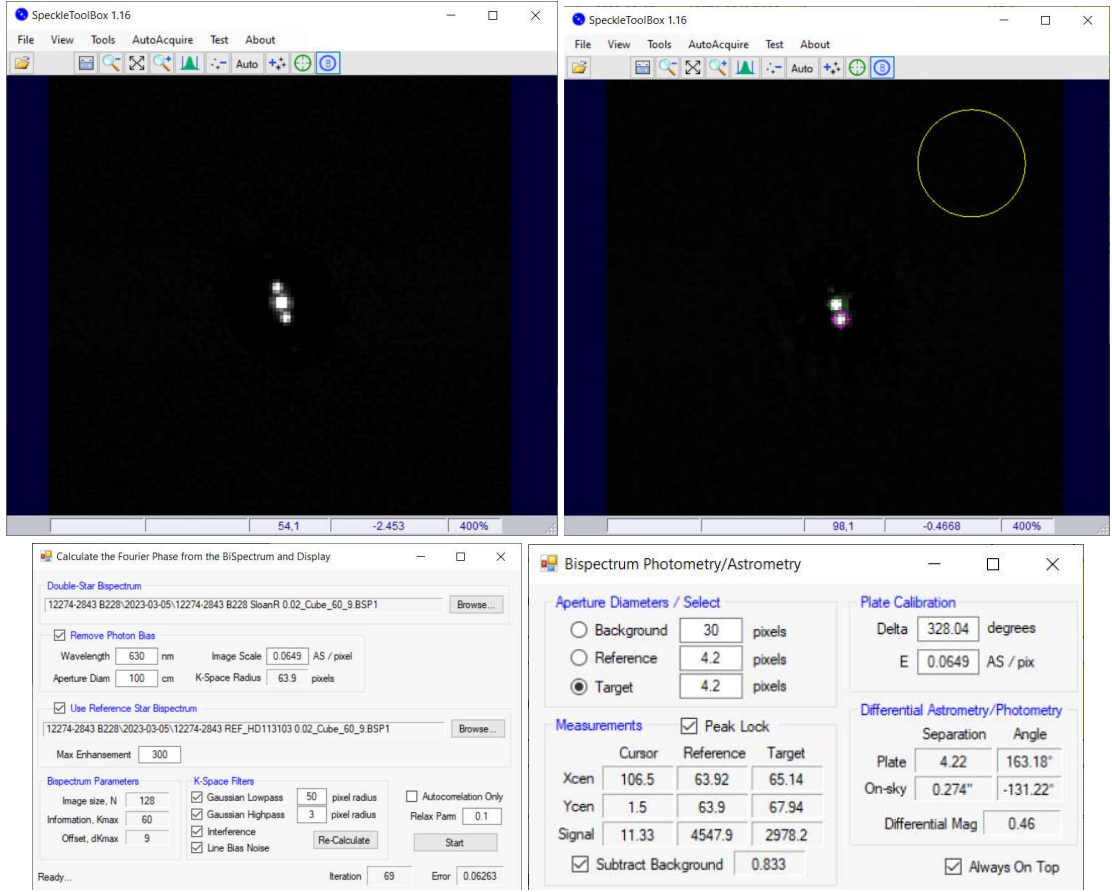

*Figure 5: Steps of STB bispectrum data reduction for 0.020-seccond exposure images of the close binary B228. Upper left: the autocorrelation is the starting point for bispectrum processing. Lower left: this window provides user control and input, as well as monitoring of iteration convergence during phase recovery. Upper right: the result of bispectrum convergence. The false lobe of the autocorrelation has disappeared, and the two real star images now have both the correct positions and correct relative brightness. Lower right: the measurement window provides inputs for calibration and measurement apertures, yielding astrometry and photometry results in the lower right portion. The primary, secondary and background apertures seen above were manually centered on the image, guided by star centroid indicators when "peak lock" was checked.*

Astrometric measurements of the image are done with input parameters from the plate solution (scale and orientation), an adjustable aperture for each star, and a background aperture. These three apertures are used instead of the traditional annuli of conventional photometry because two stars are measured simultaneously. Placement of apertures on the stars is aided by an "X" indicating the aperture centroid of each star. The average background level within the background aperture is subtracted from each star aperture for measurement of brightness.

Recovery of the wavefront phase by bispectrum analysis results in each star having the correct flux, therefore photometry can be done on the diffraction-limited image, resulting in the correct delta magnitude between the binary components. For photometry, the same apertures are used as for astrometry. However, the same diameter must be used for both stars because they both have the same point spread function; the primary star appears larger only because it rises higher above the background noise level. Table 2 summarizes the astrometry and delta magnitude results.

*Table 2: Initial bispectrum measurements of WDS 12274-2843 B228 with the 1-meter speckle system. θ, ρ and Δ Mag are Position Angle, Separation, and Magnitude difference of the two component stars, respectively. Exposures were 20 milli-seconds in the Baader SDSS r′ filter band (center WL = 628nm, FWHM = 135nm). HD 113103 was the Reference star. These measurements are by only one observer (RW).*

| WDS 12274-2843 B228 | | | | |
|---|---|---|---|---|
| Date | # Frames | θ Obs (deg) | ρ Obs (") | Δ Mag |
| 2023-03-05 2023.175 | 263 | 131.22 | 0.274 | 0.46 |
| 2023-03-08 2023.183 | 500 | 128.44 | 0.272 | 0.51 |
| 2023-03-09 2023.186 | 500 | 131.52 | 0.279 | 0.62 |
| Overall Average ⇒ | | 130.39 | 0.275 | 0.53 |
| Overall Std Deviation ⇒ | | 1.70 | 0.004 | 0.08 |

Shown below in Figure 6 are the final 128x128 Bispectrum images from the March 2023 run on WDS 12274-2843 B228.

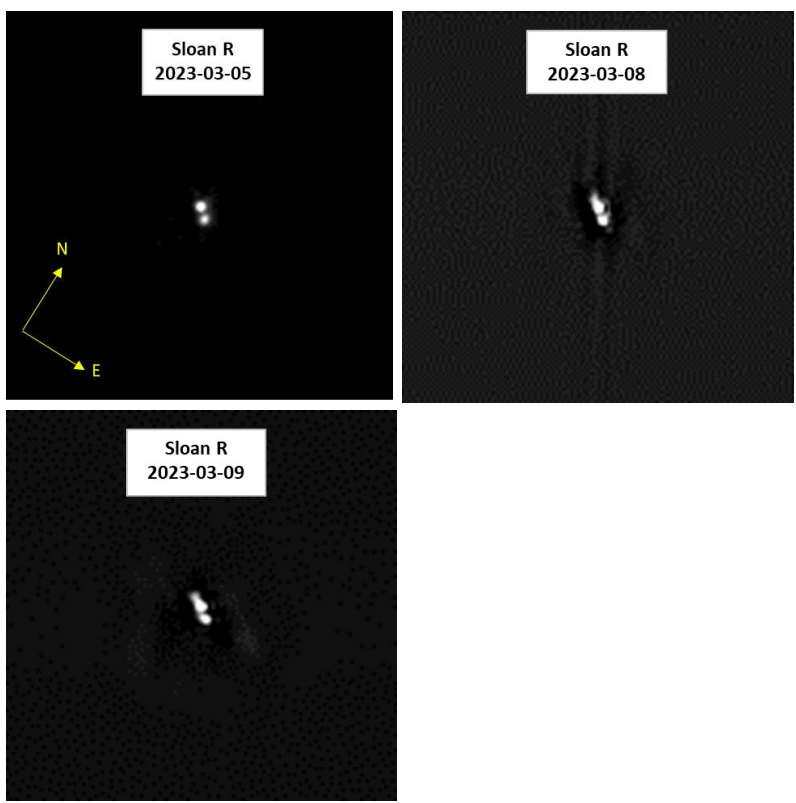

*Figure 6: Images resulting from bispectrum analysis of WDS 12274-2843 B228 on each of three nights. Image orientation on the sky, based on the plate solution, is shown for the first night. It did not change because the telescope de-rotator compensates for alt-azimuth field rotation, and the camera was not touched during the observing run.*

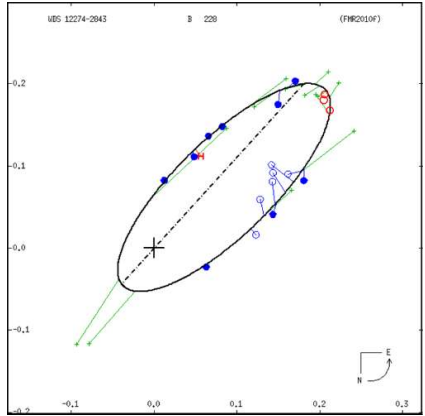

*Figure 7. The WDS orbit plot for 12274-2843 B228, with the three new observations added as open red circles.*

The WDS orbit is shown in Figure 7, where green + symbols are micrometer, open blue circles are the visual interferometer technique, and solid blue circles are speckle. The measurements from Table 2 have been added to the plot as open red circles (Buchheim 2017). There are significant differences in θ over three nights (standard deviation = 1.7 degrees from Table 2). This is likely because the separation is approaching the Rayleigh criterion for a 1-meter aperture, but the points are still grouped near the predicted orbit.

## 4. Technical Explorations

Besides employing our group's usual reduction process with the Speckle Tool Box, we explored four technical areas as described below.

*Double Gaussian Algorithm*

In the final bispectrum analysis with the Speckle Tool Box, once the bispectrum image has been reconstructed, diaphragms of appropriate size are manually centered over the binary's primary and secondary stars. Although the diaphragms are positioned manually, a 'Lock' feature that locks on to the calculated centroids of the two stars works well to give totally repeatable results if the two stars are cleanly separated and bright enough. However, if the stellar images are somewhat merged or too faint, then it is necessary to manually place the diaphragms, using human visual judgement for their placement. This is a difficult process when one is dealing with fractional pixel values with star images just a few pixels across.

One of us (James Armstrong) developed an algorithmic final reduction process that fits two 2D Gaussian curves (that approximate the point spread function) to the .fits image of the final Bispetrum Analysis image. The counts in each pixel are assumed to approximate the functional form:

$$I = A_1 * e^{-((x-x_1)/\alpha_{x1})^2 - ((y-y_1)/\alpha_{y1})^2} + A_2 * e^{-((x-x_2)/\alpha_{x2})^2 - ((y-y_2)/\alpha_{y2})^2}$$

Where I is the number of counts, A is a fit amplitude of each Gaussian, x is the x coordinate of the pixel and $\alpha^2$ is the variance. The centroids of the stars are taken to be located at $x_1$, $y_1$ and $x_2$, $y_2$, for stars 1 & 2.

The routine is written in Python and uses the curve-fit routine from the SciPy package. The curve-fit routine performs a nonlinear least squared fit to the function. It should be noted, for anyone who wishes to follow this work themselves, that curve-fit only performs fitting to 1D functions. To achieve 2D fitting, the function should be "raveled" or turned into a 1D function by reshaping the 2D X and Y coordinate arrays, and the image to be fit into a 1D array arrays.

One important feature of the fitting routine is that an initial guess is still required. Currently, the initial guess is provided from a manual estimate of the centroids of each star. For well resolved stars (centroids separated by more than 1 full-width-at-half-max, FWHM), the routine is not particularly sensitive to the initial guess and, in some test cases, the routine has performed well even with guesses that were intended for different images.

*SURFACE Reduction*

Florent Losse's REDUC program (Losse 2023) is a widely used method for double star measurements. REDUC incorporates an analysis method, SURFACE (Morelet & Salaman 2005) for analysis of close binaries where the light of the stars is overlapping. SURFACE uses an empirical equation proposed by Pierre Bacchus, of the mathematical form $f(x) = [\exp(-ax^2)] / [1 + bx^2]$, to model the illumination decrease of the point spread function from its center (Agati et al. 2009). Use of the SURFACE utility is clearly explained in the REDUC help tab. Bispectrum images from STB (Figure 6) were modeled to obtain the measurements for B228 shown in Table 3. Delta magnitude results of SURFACE are given to only one decimal place, but modeled brightness values are provided for each star so delta magnitude could also be calculated. SURFACE is an iterative program that starts with an initial guess by the user and converges toward a solution; the number of iterations is included in Table 3. The average ρ is almost 5% larger than that of Table 2, but average θ is less than 1 degree different. The Math Image option gives a graphical representation of the Surface solution, as seen in Figure 8.

*Table 3: Surface results for B228. The Angle and Scale are calibrations from the independent plate solution for each night, while the position angle (θ), separation (ρ), and delta magnitude (Δ Mag) are from converged solutions of SURFACE. The number of iterations for convergence is at right.*

| Date | Angle | Scale | θ (°) | ρ (″) | Δ Mag | Iterations |
|---|---|---|---|---|---|---|
| 2023-03-05 | 328.04 | 0.0649 | 132.75 | 0.287 | 0.4 | 5 |
| 2023-03-08 | 328.91 | 0.0649 | 131.28 | 0.285 | 0.5 | 8 |
| 2023-03-09 | 328.90 | 0.0649 | 128.99 | 0.291 | 0.6 | 6 |
| Average | | | 131.01 | 0.288 | | |
| Std Dev | | | 1.89 | 0.003 | | |

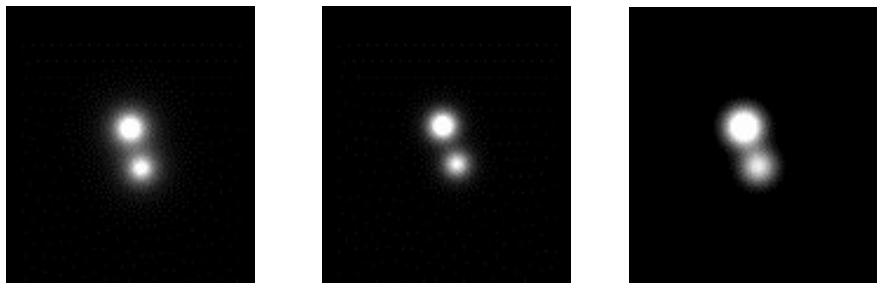

*Figure 8: SURFACE models of B228 for the nights of 2023-03-05, -08, and -09 (left to right).*

*Repeatability of Camera Plate Scale and Rotation Angle*

For speckle interferometry, it is best to adjust the plate scale of the optical system (via a Barlow or similar magnification) to provide the best possible speckle images—typically around 5-10 pixels across the first Airy disk. Depending on the camera, the resulting field-of-view may be to small too have a sufficient number of stars in it to obtain the plate solution needed to derive the accurate plate scale and camera angle for interpretation of the speckle images.

There are several solutions to this problem. For instance, one could obtain a plate solution in a crowded field and then slew to the sparse field that contains the target. This requires, however, confidence that the camera angle and plate scale will not change when slewing from one area of the sky to another.

In February 2023, a series of speckle observations were obtained by one of us (Scott Dixon) with the 1.0-m telescope of double stars in a region of the sky bounded by 6 to 8 hours RA and +5 to -5 degrees of Dec. Each speckle observation was accompanied by a 60 second exposure image to allow determination of the camera rotation angle on the sky by plate solution. For each of these observations the desired camera rotation angle position was set to a fixed value. Successful plate solutions using PS 3.80 were obtained for 17 observations on 5 nights. The resulting scales and field rotation angles resulting from these solutions are shown in Table 4 below. The result of these plate solutions indicates the camera angle setting was repeatable to 0.55 degrees 1 sigma over all 5 nights. However, on individual nights the standard deviation (except the first image on the first night) was 0.1 degrees or less.

We concluded that:

(1) the plate scale is very solid around 0.065 arcseconds/pixel; any variation from this is in the noise

(2) the camera angle within a night varies about 0.10 degrees, also down in the noise. However, the night-to-night variation of the camera angles is larger and may sometimes be significant. Therefore, if there isn't a valid plate solution for a specific speckle observation it is safe to use 0.065 arcsec/pixel as the plate scale and the mean camera angle for that night or the average of good plate solutions excluding outliers. Using an across-nights mean camera angle could introduce significant camera angle error.

Table 4: Plate solutions from the 1.0-meter telescope in February 2023.

| WDS Target | Date | Scale | Angle |
|---|---|---|---|
| 07411-0124 | 2023-02-03 | 0.065 | 275.185 |
| 07527+0323 | 2023-02-03 | 0.065 | 274.102 |
| 07573+0108 | 2023-02-03 | 0.065 | 274.137 |
| 08024+0409 | 2023-02-03 | 0.065 | 274.134 |
| | | | |
| 06173+0506 | 2023-02-05 | 0.065 | 273.826 |
| 06293-0248 | 2023-02-05 | 0.065 | 273.803 |
| | | | |
| 06173+0506 | 2023-02-06 | 0.065 | 275.140 |
| 06478+0020 | 2023-02-06 | 0.065 | 275.150 |
| 07527+0323 | 2023-02-06 | 0.065 | 275.137 |
| 07546-0248 | 2023-02-06 | 0.065 | 275.179 |
| 08024+0409 | 2023-02-06 | 0.065 | 275.170 |
| 08024+0409 | 2023-02-06 | 0.065 | 275.143 |
| | | | |
| 06173+0506 | 2023-02-07 | 0.065 | 275.122 |

| | | |
|---|---|---|
| 06478+0020__2023-02-07 | 0.065 | 275.154 |
| 06575+0253__2023-02-07 | 0.065 | 275.186 |
| 06575+0253__2023-02-07 | 0.065 | 274.957 |
| 07546-0248__2023-02-09 | 0.065 | 328.961 |

*Reduced Regions-of-Interest*

While initially reducing some of the observations, one of us (Mark Harris) noted that, at least for the closest binaries (such as WDS 04460-6605) the image quality of the final reductions were enhanced if a smaller region-of-interest (ROI) was used than the initial speckle images of 512x512 pixels. The rather dramatic improvement in image quality due to reduced ROI are shown in Figure 9.

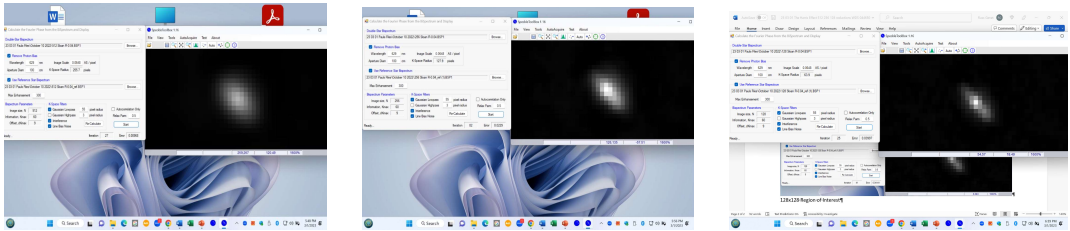

*Figure 9: Bispectrum Analysis reduction of WDS 04460-6605 using three different regions-or-interest. Left to right: 512x512, 256x256, and 128x128 pixels.*

## 5. Conclusions

Overall conclusions about the performance limitations of the 1.0-meter telescope and the precision and accuracy of automated speckle interferometry measurements with this telescope will have to wait until the large number of binaries we have already observed but have not reduced yet are analyzed. We can, however, draw a few preliminary conclusions based on our observations of WDS 04460-6605.

- While further refinements in the automated speckle interferometry process will undoubtedly improve observational efficiency and ease of use, we conclude that automated speckle interferometry can provide useful observational results.

- Human-induced variance during the manual reduction process can be significant for close binaries where the "Lock" feature of the Speckle Tool Box cannot be used.

- However, the use of algorithmic two-Gaussian curve fitting processes such as SURFACE and Armstrong's method may eliminate this variance, although their accuracy is yet to be determined.

- Reducing the region-of-interest during the reduction process can improve results, although this reduction may be limited by the separation of the two stars as well as the tracking accuracy of the telescope.

- Finally, the plate (pixel) scale was found to be remarkably constant. However, the camera angle, while quite constant over the course of a night, was found to vary significantly between nights.

## Acknowledgements

We thank Mike Selby for providing observing time on the 1.0-meter telescope, El Sauce Observatory for maintaining the telescope, and the United States Naval Observatory for providing information on past observations of WDS 04460-6605, and Robert Buchheim and Richard Harshaw for external reviews.